\documentclass{ws-mpla}

\RequirePackage{lineno}
\usepackage{color}
\usepackage{xspace}

\makeatletter
  \newcommand\@ptsize{1}
\makeatother
\usepackage{setspace}

\newcommand{\ttbar}{\ensuremath{t\overline{t}}}
\newcommand{\ljets}{\mbox{$\ell$+jets}}
\newcommand{\dil}{\mbox{$\ell\ell$}}
\newcommand{\met}    {\mbox{$\not\!\!E_T$}}
\newcommand{\ppbar}{\ensuremath{p\bar{p}}}
\newcommand{\qqbar}{\ensuremath{q\bar{q}}}
\newcommand{\ifb}{\mbox{fb$^{-1}$}}
\newcommand{\pt}{\mbox{$p_T$}}

\newcommand{\qbar}{\ensuremath{\overline{q}}}
\newcommand{\tbar}{\ensuremath{\overline{t}}}
\newcommand{\pbar}{\ensuremath{\overline{p}}}

\newcommand{\commentOut}[1]{}

\begin{document}

\markboth{Fr\'ed\'eric D\'eliot, Yvonne Peters, Veronica Sorin}
{Top Quark Physics at the Tevatron} 

\catchline{}{}{}{}{}

\title{Top Quark Physics at the Tevatron}

\author{Fr\'ed\'eric D\'eliot}

\address{Institut de recherche sur les lois fondamentales de l'univers, Service de physique des particules\\
CEA-Saclay, bat 141, F-91191 Gif-sur-Yvette Cedex, France\\
frederic.deliot@cea.fr}

\author{Yvonne Peters\footnote{also at DESY.} \footnote{Corresponding author.}}

\address{University of G\"ottingen\\
Friedrich-Hund-Platz 1\\
37077 G\"ottingen, Germany\\
reinhild.peters@cern.ch
}

\author{Veronica Sorin}

\address{Institut de Fisica d'Altes Energies, Universitat Autonoma de Barcelona \\
E-08193 Bellaterra, Barcelona, Spain\\
vsorin@ifae.es}

\maketitle

\pub{Received (6 February 2013)}{Revised (Day Month Year)}

\begin{abstract}
The heaviest known elementary particle, the top quark,
 was discovered in 1995 by the CDF and D0 collaborations at the
  Tevatron proton-antiproton collider at Fermilab. Since its discovery, a large program was set in motion by the CDF and D0 collaborations to characterize the production and decay properties of top quarks, and investigate their potential for searches of new phenomena beyond the standard model.
During the past 20 years,  new methods were developed and implemented to improve the measurements and searches for new physics in the top-quark sector.  This article reviews the achievements and results obtained through studies of the top quark at the Tevatron. 
\end{abstract}

\ccode{PACS Nos.: 14.65.Ha}


\section{Introduction}
\label{sec:Introduction}


 The top quark ($t$)  was discovered in  
1995 by the CDF and D0 collaborations at the Tevatron proton antiproton ($p\bar{p}$) collider at
Fermilab.\cite{cdftopdiscovery,d0topdiscovery} With a mass of  $m_t=173.2 \pm 0.9$~GeV,\cite{topmassworldaverage} 
the top quark is the heaviest known elementary particle, which has led to speculations that it may
play a special role in the mechanism of
electroweak symmetry breaking. 
The lifetime of the top quarks is shorter than their hadronization time, which provides a unique 
opportunity to study the properties of essentially bare quarks. 
Since its discovery, intensive programs have been undertaken at the CDF and D0
experiments, and recently at the  LHC, to categorize the properties of the newly 
discovered particle, believed to be the top quark of the standard model of particle physics (SM).
In particular, the production and properties 
are studied and compared to predictions from SM calculations,
and searches for physics beyond the SM (BSM) in the top-quark sector have been performed 
through comparisons of decay modes and departures from expectations in different final states. 
This review presents  the status of our understanding of the top
quark that has been  gained from experiments at the Tevatron.

\subsection{A brief History of the Top Quark}

One of the greatest legacies of the Tevatron is the observation of the top 
quark. The existence of the top quark was predicted years before its
discovery.\cite{TopHistory,toplegacyyvonne} After the discovery of the  $\tau$-lepton
at SLAC in 1976, and the  upsilon, and thereby the $b$-quark, in 1977
at Fermilab, the fermion sector of the SM had to be
extended from two to three generations. Since the SM predicted a doublet
for each quark generation, the up-type partner of the 
$b$-quark, i.e., the top quark, was missing. Based on this realization,
a race was launched to find the predicted top quark, first, in searches for
$t\bar{t}$ bound states, as, for example, at the $e^{+}e^{-}$ colliders PETRA at
DESY and at Tristan at KEK. Next, the top quarks were sought in $W$-boson decays ($W \rightarrow tb$) at the UA1 and
UA2 experiments at CERN. In 1988, CDF joined the race to find the top
quark in the initial running phase of the Tevatron, and D0 joined CDF for Run~I in 1992. 
(Run~I lasted from 1992 to 1996, colliding $p$ and $\bar{p}$ at a center-of-mass energy of 1.8~TeV, 
and collecting about 100~pb$^{-1}$ of data per experiment.)
Due to different
designs of the CDF and D0 detectors  during Run~I, the experiments
used different strategies in their searches for the top quark in the $p\bar{p} \rightarrow t\bar{t}$ final state.  
CDF focused on $b$-jet identification to reduce background, and D0 on the use of topological
information. The first searches at the Tevatron set lower limits on
the mass of the top quark that were above the  mass of the $W$ boson, eliminating the possibility of observing the 
top quark  via decays of $W$-bosons.  It took until 1994 however to find first evidence for
possible $t\bar{t}$ production\cite{cdftopevidence1,cdftopevidence2}. 
Definitive evidence for the observation of the top quarks was 
presented on  February 24th, 1995, when CDF and D0
simultaneously submitted their results for publication.\cite{cdftopdiscovery,d0topdiscovery}
The  observations were based on 50~pb$^{-1}$ at D0, and 67~pb$^{-1}$ at CDF.
After the first observation of top quarks in \ttbar\ events, it took 14 more years until the electroweak (EW) 
production of single  top quarks was reported by the CDF and D0 
collaborations\cite{cdfsingletop1,cdfsingletop2,d0singletop} during Run~II, which lasted from 2001 until  2011, with 
data taken at a $p\bar{p}$ center-of-mass energy of 1.96~TeV, recording about 10~fb$^{-1}$ of integrated luminosity 
per experiment.

\section{Production and Decay in the Standard Model}
\label{sec:SM}


At hadron colliders, top quarks are expected to be produced mainly as \ttbar\ pairs through 
the quantum-chromodynamic (QCD) strong interaction, with quark-antiquark
annihilation ($\qqbar \to \ttbar$) and gluon fusion ($gg \to \ttbar$) corresponding to the primary processes. 
At the Tevatron, $\qqbar$ is responsible for $\approx 85\%$ of the total \ttbar\ production cross section.
The latest calculations, at next-to-next-to-leading order (NNLO) in perturbative QCD for the 
$\qqbar \to \ttbar$ component, and at approximate NNLO for $gg \to \ttbar$, both
including soft-gluon resummations to next-to-next-to-leading logarithmic (NNLL) accuracy,
yield a total cross section of 
$\sigma(\ppbar \to \ttbar) = 7.24^{+0.23}_{-0.27}$~pb for a top-quark mass of $m_t=172.5$~GeV.\cite{SMtheory_NNLO}
The second contribution to the production of top quarks takes place through EW processes 
that involve production of single top quarks. These final states have contributions from 
(i) the s-channel where a virtual $W$ boson mediates the $\qqbar$ annihilation 
to produce a top and a $b$~quark ($tb$), (ii) the t-channel, where a $W$ boson is exchanged in the
t-channel of an incident $qg$ system, leaving a light quark, a top quark, and a possible 
accompanying $b$~quark, in the final state ($tqb$), and (iii) the Wt-channel ($Wt$), where the 
final state consists of a $W$~boson and a top quark that arise from an initial $bg$ system, where
the $b$~quark corresponds to an intrinsic component of the proton (or antiproton) substructure ($b\bar{b}$ sea). 
Approximate NNNLO calculations for the sum of these three contributions predict a single top-quark production
cross section of $\sigma(\ppbar \to tb/tqb/Wt) = 3.58 \pm 0.14$~pb,
assuming a top-quark mass of $m_t=172.5$~GeV.\cite{Kidonakis:2006bu}  

Within the SM, the top quark decays almost 100\% of the time into a $W$~boson and a $b$~quark, and
signatures for \ttbar\ events can therefore be classified according to the decays of the $W$ bosons. 
When both $W$~bosons decay to $q' \bar{q}$ pairs, the final states of \ttbar\ are referred as the alljets channel. 
It has a large branching ratio, but it is also contaminated by a large background from generic multijet production.
The dilepton+jets channel (\dil) corresponds to leptonic decays of both $W$~bosons, either into $e \nu_e$ or $\mu \nu_{\mu}$
(with contributions from leptonic decays of the $\tau$-lepton: $\tau_{\ell} \to \ell \nu_{\ell} \nu_{\tau}$). 
This channel has a small branching ratio, but also very little background contamination from, e. g., $\ell^+ \ell^-$ Drell-Yan
($q \bar{q} \to \ell^+ \ell^-$) or from $WW, ZZ$ or $WZ$ (diboson) production. 
The lepton+jets channel (\ljets) consists of events where one $W$ boson decays leptonically into
$e \nu_e$ or $\mu \nu_{\mu}$ and the other $W$ boson into  $q' \bar{q}$.
This \ttbar\ final state has a large branching fraction, and a manageable background, 
mainly from $W$+jets and multijet production.
Channels with a $\tau$-lepton decaying to hadrons+$\nu_{\tau}$ ($\tau_h$) are 
treated separately, and have significant background from light jets misidentified as hadrons from $\tau$ decay.  
For some analyses, no explicit lepton-identification criteria are required to maintain sensitivity to all
leptonic $W$-boson decays. However, in such cases, a large imbalance in transverse momentum (\met), arising from undetected 
energetic neutrinos, is expected to be present in association with the jets. These criteria define the \met+jets channel.
CDF and D0 have measured the \ttbar\ production cross sections in almost all of the above \ttbar\ decay modes.
This provides the possibility of comparing measurements with predictions from the SM
among the individual channels, and check thereby for any indication of physics beyond the SM. 

Because of significant background, single top-quark production is studied only in final states where
the $W$~boson decays leptonically.
The main backgrounds to EW single top-quark measurements arise from \ttbar, $W$+jets and multijet production.  

The following sections present an overview of the top-quark measurements performed at the Tevatron.
Section~3 provides results on the measurements of cross sections as well as their
ratios in different channels. 
Properties of top quark are examined in Sec.~4. This starts with the measurements of $m_t$, followed
by issues pertaining to helicities of the $W$ bosons in $t \to Wb$ decays, angular production properties
of $t$ and $\bar{t}$ quarks, and, finally, measurements limited by statistical uncertainties that involve
correlations in electric charges, in $t$ and $\bar{t}$ spin and color flow in \ttbar\ events, as well as issues
pertaining to the width of the top quark and anomalous couplings examined through
combined \ttbar\ and single top-quark analyses.
Section~5 describes searches for new phenomena in the top-quark sector, and a brief overall summary is given 
in Sec.~6.

\section{Cross Sections for the Production of Top Quarks}
\label{sec:prod}

First  measurements of the \ttbar\ production cross section were made at the Tevatron 
using the small data sample collected during Run~I. The integrated luminosities in Run~II were a factor 
of $\approx 100$ larger, 
and led to improvements in techniques to separate signal from background, and thereby  to a reduction in systematic uncertainties. As a result, 
measurements of the cross section improved rapidly in precision, reaching current uncertainties of $<10\%$ that provide stringent checks of QCD predictions.\cite{SMtheory_NNLO,SMothers1,SMothers2,SMothers3}  
The next sections describe the present status of the cross-section measurements at the Tevatron, and outline 
the analyses techniques utilized in these measurements.

\subsection{From Discovery to Precision of Cross Sections}
While at the beginning of Run~II,  measurements of \ttbar\ production cross sections had uncertainties 
of  $\approx 30\%$,\cite{RunIttbar} nowadays, single measurements have reached precisions as small as 7\%.\cite{cdfljtopo} The most recent 
combination of CDF and D0 results for the total \ttbar\ cross section, based on 8.8~\ifb\ of data, is  $\sigma(\ppbar \to \ttbar) = 7.65 \pm 0.42 $~pb  (for  $m_t=172.5$~GeV),
corresponding to a precision of 5.5\%.\cite{tevxscombo} Because BSM processes can affect differently each of the 
\ttbar\ decay modes, measurements at the Tevatron experiments have sampled many final states. 
Event signatures are usually characterized by the presence of leptons and jets of large transverse momentum (\pt),
and often of large \met. Both collaborations have developed techniques that are best suited to exploit the 
strengths of each detector for the reconstruction and identification of 
such observables.\cite{cdfdetector,d0detector1,d0detector2}


Jets are reconstructed using different implementations  of cone algorithms,\cite{cdfd0jets} with possible 
requirements of having one or two jets that are candidates for arising from the evolution of bottom quarks.
Such inference is based on methods of $b$-tagging that take advantage of the long lifetimes of hadrons that contain $b$ quarks, as indicated by the presence of displaced  ("secondary") vertices within jets that correspond to 
decays of $b$-quarks.\cite{secv} The use of $b$-tagging techniques has provided \ttbar\ samples of sufficient purity to calculate a cross section just by  counting events.  Multivariate approaches  have also been developed
to distinguish jet flavors by combining properties of tracks and displaced vertices associated with jets.\cite{btagNN} 

Another technique utilized for measuring the \ttbar\ cross section consists of fitting a discriminant 
variable constructed from topological or kinematic information  to distinguish signal from background 
processes.\cite{cdfljtopo,d0lj} This method makes no assumptions about the flavor of jets in \ttbar\ events, but 
exploits differences in variables such as the magnitude of the sum of absolute \pt\ values of objects in an
event ($H_T$), or relies on differences in event topologies that are expected for signal and background.

The development of multivariate techniques has played 
a particularly important role at the Tevatron in the search for single top-quark production. Although its cross section is  
about half that of the \ttbar\  process, the background contribution from $W$+jets production is overwhelming. With only one top quark 
in the final state, the signature for single top-quark events is not as restrictive in reducing background as 
for \ttbar\ production. To observe EW production of single top quarks requires therefore more inventive use 
of kinematic properties and $b$-tagging to define a discriminant that is sensitive to
regions of phase space corresponding to  large signal relative to background. 
Such regions of data can be fitted successfully to contributions from signal and 
background,  in a procedure that takes into account  uncertainties related to the normalization as 
well as the dependence on modeling of differential distributions in any given  input variable. This 
method was applied by both CDF and D0 collaborations in their observation of a single top-quark signal,\cite{cdfsingletop1,cdfsingletop2,d0singletop} and in the measurement of the cross section and the 
CKM matrix element $|V_{tb}|$.\cite{ckm1,ckm2} Recent work has been extended to provide essentially model-independent 
results in the $t$-channel,\cite{d0tch} and to searches for new physics, to be discussed in Sec.~5.



\subsection{Current Status of cross sections}
The most precise measurements of cross sections are obtained using the \ljets\ channels. 
CDF has complementary analyses, one based on counting events in samples with small background contributions,
which is achieved by requiring the 
presence of two $b$-jets in each event, and, another, through a topological approach in the fitting of the output of an artificial neural network (NN)
to events with  more than two jets. In both types of measurements, the largest systematic uncertainty, which is due to the uncertainty 
on the integrated luminosity, can be reduced by normalizing to the observed inclusive yield of  Z bosons in two-lepton final states, which can be calculated with great reliability.\cite{cdfljtopo}  Results based on 4.6~\ifb\ of data are combined using a best linear unbiased estimate (BLUE),\cite{BLUE-method-2,BLUE-method-3} 
that yields a value of $\sigma(\ppbar \to \ttbar) = 7.32 \pm 0.52 {\rm (stat+syst)}$~pb, assuming a 
mass of $m_t=172.5$~GeV. 
The D0 collaboration finds a cross section of  $\sigma(\ppbar \to \ttbar) = 7.78^{+0.77}_{-0.64}{\rm (stat+syst)}$~pb, by
combining channels and using measurements that exploit both kinematic information and $b$-tagging information 
in 5.3~\ifb\ of data.\cite{d0lj}

As mentioned before, although the \dil\ channel of \ttbar\ production has a smaller branching fraction than the  \ljets\  mode, it has the advantage of a better  
signal to background ratio, even without use of $b$-tagging. These pure samples of \ttbar\ events have also been explored by both 
collaborations. Using a sample  of 5.4~\ifb, D0 measures a cross section of $\sigma(\ppbar \to \ttbar) = 7.36^{+0.90}_{-0.79}{\rm (stat+syst)}$~pb 
for a top-quark mass of $m_t=172.5$~GeV.\cite{d0dil} The cross section is extracted through a fit of the 
output of a NN $b$-tagging algorithm to data. At CDF, a counting experiment  yields 
$\sigma(\ppbar \to \ttbar) = 7.66 \pm 0.46{\rm (stat)} \pm 0.66{\rm (syst)} \pm 0.47{\rm (lumi)}$~pb for a data 
sample of 2.8~\ifb.\cite{cdfdil} 
%

Identifying leptons, in particular  electrons and muons, leads to drastically reduced backgrounds and therefore 
to simpler selection requirements for \ttbar\ measurements. Nevertheless, the large sets of data, from well understood 
detectors, and the advances in analysis techniques have helped to produce excellent measurements even in the alljets channel. In this mode, 
final states are characterized by at least six jets, a signature that is similar to that of generic multijet production. To distinguish  \ttbar\ events from multijet background, $b$-tagging, kinematics and topological information is combined in a variety of ways. 
CDF utilizes the output of a likelihood fit exploited to measure $m_t$ in a sample 
of events defined by a NN-based kinematic selection, together with the requirement of having at least 1 $b$-tagged jet, to extract a 
cross section of $\sigma(\ppbar \to \ttbar) = 7.2 \pm 0.5{\rm (stat)} \pm 1.1{\rm (syst)} \pm 0.4{\rm (lumi)}$~pb 
for a top-quark mass of $m_t=172.5$~GeV in 2.9~\ifb\, of data.\cite{cdfallj}  D0 reduces the contamination from multijet events by requiring at least 
two $b$-tagged jets, and by implementing  fits to the cross section based on a likelihood discriminant constructed from kinematic and 
topological information.\cite{d0allj}  For a data sample of  1~\ifb, D0 finds $\sigma(\ppbar \to \ttbar) = 6.9 \pm 1.3{\rm (stat)} \pm 1.4{\rm (syst)} \pm 0.4{\rm (lumi)}$~pb, for $m_t=175$~GeV.

These techniques have been extended to measurements in modes that have $\tau_h$ decays with 
signatures corresponding to the presence of large \met\ and at least four jets.  To identify the $\tau_h$ decay products, which appear as narrow jets of hadrons, a set of neural networks is used by the D0 collaboration,\cite{d0tl1,d0tl2} and a two-cones algorithm (a signal cone and an isolation annulus)\cite{2cones}  by CDF. 
The large background from multijet events is reduced through a restrictive cutoff on the output of a NN in the CDF analysis (that is also used to  extract a value of  $m_t$), as well as through differentiating background 
from signal in a fit of a distribution in NN output to the data, as  done by D0. 
Both experiments apply their methodologies to samples with at least one $b$-tagged jet. Using 1~\ifb\ 
of data, D0 measures a cross section of 
$\sigma(\ppbar \to \ttbar) = 6.9 \pm 1.2{\rm (stat)}^{+0.8}_{-0.7}{\rm (syst)} \pm 0.4{\rm (lumi)} $~pb for a 
top-quark mass of $m_t=170$~GeV,\cite{d0tau} while in 2.2~\ifb\ of data, CDF finds a cross section of 
$\sigma(\ppbar \to \ttbar) = 8.8 \pm 3.3{\rm (stat)} \pm 2.2{\rm (syst)} $~pb for a top-quark mass of $m_t=172.5$~GeV.\cite{cdftau} 
CDF also measures the cross section in the \met+jets channel, where no lepton identification is required, but by vetoing 
 electrons and muons of large \pt\ to enhance the contribution from $\tau$ decays.\cite{cdfmetj} Event selection includes 
 using a NN, and the cross section is measured by counting $b$-tagged jets. Background $b$-tag rates are obtained from  three-jet data samples, which do not contain significant contributions from \ttbar\ production. 
Utilizing 2.2~\ifb, the measured cross section corresponds to 
$\sigma(\ppbar \to \ttbar) = 7.99 \pm 0.55{\rm (stat)} \pm 0.76{\rm (syst)} \pm 0.46{\rm (lumi)} $~pb for a 
top-quark mass of $m_t=172.5$~GeV.

The measured \ttbar\ cross section has become a precision standard  at the Tevatron. Results have been found 
in agreement among channels, and each experiment has provided a combination of their most precise results, 
finding at CDF a value of $\sigma(\ppbar \to \ttbar) = 7.71 \pm 0.31{\rm (stat)} \pm 0.40{\rm (syst) } $~pb, and at 
D0 a cross section of  $\sigma(\ppbar \to \ttbar) = 7.56 \pm 0.20{\rm (stat)} \pm 0.56 {\rm (syst)}  $~pb. 
These are the measurements entering the Tevatron combination\cite{tevxscombo} mentioned in the previous section,
and are in excellent agreement with expectations from the SM.

Besides  the production rate in different decay channels, 
another way to probe the presence of physics beyond the SM is by means of ratios of cross sections among different final states and  the study of differential 
 cross sections for a variety of variables. In particular, D0 has  measured the ratios of  cross sections for \dil\ channels relative to 
 \ljets\  channels ($R^{ll/lj}_{\sigma}=\sigma^{ll}_{\ttbar}/\sigma^{lj}_{\ttbar}$) and  between the $\tau$l 
channel and the \dil\ and the \ljets\ channels ($R^{\tau l/ll-lj}_{\sigma}=\sigma^{\tau l}_{\ttbar}/\sigma^{lj\&ll}_{\ttbar}$) in 1~\ifb\ of data, by defining all modes to be mutually exclusive.\cite{d0rsigma} Results for $R^{ll/lj}_{\sigma}=0.86^{+0.19}_{-0.17}$ and $R^{\tau l/ll-lj}_{\sigma}=0.97^{+0.32}_{-0.29}$ are in agreement with 
the SM expectation of unity. The presence of contributions from  BSM can  be checked through the ratio of   branching fractions  
$R_b=B(t \to Wb)/B(t \to Wq)$, where $q$ represents any possible down-type quarks ($q=d,s,b$). This measurement is performed in the \ljets\ and \dil\ channels by both CDF and D0 experiments, by fitting the number of events with 0, 1 and 2 b-tagged jets in the \ttbar\ candidate samples. \cite{d0rb,cdfrb}  In the most recent analysis by D0 , the procedure was extended to fit the $b$-tag  NN output to data in  \dil\ events, resulting in $R_b$ to be $R_b=0.90 \pm 0.04{\rm (stat+syst)}$ using 
5.4~\ifb of data.\cite{d0rbnew}. Results from both experiments were found consistent with the SM prediction of R close to unity.

%

To check  predictions from  perturbative QCD, and to define some   generic
tests of the presence of physics BSM, CDF and D0 study  differential cross sections for a variety of 
 variables in $\ttbar$ events. 
While CDF examines the $\ttbar$ invariant mass ($m_{t\bar{t}}$) in 
2.7~fb$^{-1}$,\cite{cdf_diffxsec} D0 considers the \pt\ of the  top quarks in 1~fb$^{-1}$ of
data.\cite{d0_diffxsec} For both studies, events must have at least one $b$-tagged jet in the
\ljets\ final state, which provides reconstruction
of the $\ttbar$ final state with  good resolution. Both the \pt\ and $m_{t\bar{t}}$ distributions are
unfolded to the parton level, thereby correcting the data for effects of
resolution, and acceptance. D0 compares the 
$p_T$ distribution expectations from several MC generators, while CDF calculates the
consistency of the $m_{t\bar{t}}$ distribution with expectations of the SM.

Although the main production mode for top quarks at the Tevatron is the $\ppbar \to \ttbar$ reaction, 
the rate for EW production of single top quarks is about half that of \ttbar, but as mentioned previously,
with a much smaller signal-to-background ratio. 
Observation of single top quarks was announced by CDF and D0 in 2009.\cite{cdfsingletop1,cdfsingletop2,d0singletop}  
This milestone was  achieved largely through the development of multivariate techniques that exploit the small differences  in
kinematic properties between single top quark production and  background processes.  Many analyses have 
been performed using a variety of methods, such as artificial neural networks,\cite{cdfNN} Bayesian NN 
(BNN) discriminant,\cite{BNN} discriminants based on matrix elements (ME)\cite{Abazov:2004cs}, boosted decision trees 
(BDT)\cite{BDT1,BDT2} and multivariate likelihood functions.\cite{cdfLF}
 These analyses 
are based on selections that require events containing  a lepton (electron or muon) of high \pt, significant 
\met\ and two  or more jets, one of which is $b$-tagged.  The sensitivity of the analyses is improved by 
separating events into different jet multiplicities, as well as 
according to the number of $b$-tagged jets in an event. Since the measurements are only partly correlated,  final results 
are obtained by combining the output of each analysis into a single discriminator. D0 uses a set of BNN discriminators, while CDF relies on
the  ``NEAT''\cite{neat} NN, which is later combined with the output of an orthogonal 
analysis that selects events requiring just jets and large \met\ .  Using 2.3~\ifb\ of integrated luminosity, D0 finds  a 
cross section for the combined $s$ and $t$ channels of  $\sigma_{s+t} =3.94 \pm 0.88$~pb (for $m_t=170$~GeV),\cite{d0singletop} and CDF, using a sample of 3.4~\ifb, measures 
$\sigma_{s+t} =2.3^{+0.6}_{-0.5}$~pb (for $m_t=175$~GeV).\cite{cdfsingletop1,cdfsingletop2}  
Measurements from both collaborations are
combined in a Bayesian analysis, yielding a cross section of $\sigma_{s+t} =2.76^{+0.56}_{-0.47}$~pb  (for a  $m_t=170$~GeV).\cite{tevsingletop} Since this cross section is proportional to $|V_{tb}|^2$, 
without assuming unitarity of the 3x3 CKM matrix (but assuming the dominance of $t \to Wb$ decays),  a value of the matrix element $|V_{tb}|=0.88 \pm 0.07$ is 
extracted for the combined result, with a lower limit at 95\% confidence (CL) of $|V_{tb}|>0.77$.

A new measurement from D0, using 5.4~\ifb\ of data, finds a cross section for $\sigma(\ppbar \to tqb + X)$ of $2.90 \pm 0.59$~pb (for $m_t=172.5$~GeV) for inclusive $t$-channel production,  without any assumption on the production rate for  the $s$-channel.\cite{d0tch} The values of  
$\sigma(\ppbar \to tb + X) = 0.68^{+0.38}_{-0.35}$~pb and $\sigma(\ppbar \to tqb + X) = 2.86^{+0.69}_{-0.63}$~pb, 
are found when the SM $tqb$ and $tb$ production rates are assumed, respectively in these two analyses.

 Large data samples have also been used to pursue direct searches for single top-quark production 
in channels such as $\tau$+jets at the Tevatron. 
While the \met\  + jets mode  is sensitive to $W \to \tau \nu_\tau$ decays, another possiblity to study events with sensitivity to  $W \to \tau \nu_\tau$ is by measuring events with $\tau_h$ decays. The latest study from 
D0 has focused on the contribution from $\tau_h$ decays in 4.8~\ifb\  of data.\cite{d0tausingletop} Using a BDT discriminator 
to identify $\tau_h$ decays, as well as to separate single top-quark events from background, the measurement yielded an upper limit on the single top quark cross section of 7.3~pb at a 95\% CL. Adding this channel to the D0 observation analysis provides an increase of signal acceptance of $\approx 30\%$ and  
a gain of $4\%$ in expected sensitivity.

\section{Properties of the Top Quark}
\label{sec:prop}

Measuring the properties of top quarks is essential for gauging to what extend they coincide with the predictions of the SM.
Many innovative techniques have been developed to measure these properties, and will be described briefly in the 
sections below, starting with the measurement of $m_t$.
Some of the precision results require large data samples, and combining of results from several sources, which 
we also discuss below.

\subsection{Measuring the Mass of the Top Quark}
\label{masssec}

An important property of the top quark is its mass.
The value of $m_t$ is a free parameter of the SM, which together with the mass of the 
$W$~boson ($M_W$) constrains the mass ($m_H$) of the Higgs boson ($H$) through EW
quantum corrections. Comparing this indirect prediction with a direct measurement of $m_H$
provides a stringent test of the consistency of the SM.

Prior to the data from Run~II, the value of $m_t$ was known only to an uncertainty of the 
order of 5~GeV.\cite{Abazov:2004cs}
Even with the $\approx 100$-fold increase anticipated in luminosity from Run~II, it was not expected that the precision on $m_t$
would improve greatly because of limitations from systematic uncertainties. Nevertheless,
the best single analyses have achieved precisions of 1.3~GeV while a combination of the best 
Tevatron results has an uncertainty of 0.94~GeV, corresponding to an accuracy of 0.54~\%.\cite{topmassworldaverage}
This makes the mass of the top quark the best known mass in the quark sector of the SM.
This achievement was realized largely through the introduction of the so called ``{\it in-situ}'' jet calibration
(see below),\cite{Abulencia:2005aj}
and through innovations in analyses, such as the matrix element (ME) approach, first used successfully
in Ref.\cite{Abazov:2004cs}. Both developments will be described below, as well as other methods of
analysis pioneered at the Tevatron for measuring $m_t$.  

Although the mass of the top quark is reflected in the kinematic distributions of all of its decay
products, there are primarily three main methods that have evolved for measuring $m_t$. These are: (i) the template method,
(ii) the ME method, and (iii) the ideogram method.
All these rely on the calibration of the measured $m_t$ through Monte Carlo (MC) pseudo-experiments, the analyses
of which is used to correct for simplifications or other assumptions of each method.
Each pseudo-experiment contains a mixture of \ttbar\ signal (simulated using MC generators) and background events
(either from MC or based on other data) that reflects the composition of the analysis samples in data.
The relation between the mean value of $m_t$ extracted in the pseudo-experiment and the 
input is usually fitted to a linear dependence, and used to  
correct the value of $m_t$ found for data. Pseudo-experiments are also used to calibrate the statistical uncertainty of a 
given method.
For \ljets\ or alljets channels, where at least one of the $W$ bosons from the top quarks decays into two 
light quarks,
the jets from the $W$ boson can be used to recalibrate the jet energy scale (JES) through an {\it in-situ} jet calibration.
For true \ttbar\ events, the invariant mass of the two jets from the $W$ boson can be constrained to the world average 
value of $M_W$,\cite{pdg} and the result used
to adjust the energy scale of all jets (JES). 
This procedure reduces the impact of the uncertainty on the absolute jet energy
in the measurement of $m_t$.

The simplest method for measuring $m_t$ is the template method, which relies on comparing the properties of
an observable whose value is
correlated with the mass of the top quark, which we wish to extract from the data. It is based on
MC distributions (templates) in this observable, generated for different values of $m_t$.
The observable found to be most strongly correlated with $m_t$ is, not surprisingly, the invariant mass of the top quark 
reconstructed from its decay
products. This reconstruction can be performed using a kinematic fit to the candidate events, assuming the constraints
of the \ttbar\ hypothesis.
To improve the statistical power of the analysis, all or the most likely solutions to the jet permutations in the 
\ttbar\ hypotheses are considered in the analysis. Often, just the reconstructed $m_t$ with the best and second-best
value of $\chi^2$ fit probability are retained.
Observables other than the reconstructed $m_t$ can also be exploited to minimize sensitivity 
to some specific uncertainty, such as the uncertainty from jet energy corrections. 
The latter was done by CDF through the use of the observed decay lengths of $B$ hadrons evolved from $b$~quarks. 
This is usually implemented in the plane transverse to the incident proton and antiproton beams.\cite{Aaltonen:2009hd}
Alternatively, variables such as the transverse momenta of leptons from decays of $W$ bosons can be used in these 
studies.\cite{Aaltonen:2009zi}
The sensitivity of the template method is not as high as that of other methods that rely on more
kinematic information in an event, or that assign larger weight to more well-measured and more likely \ttbar\ events.
However, with the large data samples available at the end of Run~II and smaller impact of statistical uncertainties, 
this method is now very competitive.
The simpler template method has therefore been used to measure $m_t$ in most \ttbar\
decay channels, as summarized in what follows.

With 5.6~\ifb\ of data, CDF performs a simultaneous fit to the \ljets\ and \dil\ channels, using {\it in-situ} jet 
calibration for events with one or two $b$-tagged jets in \ljets\ events, and for \dil\ 
data that contain either untagged or $b$-tagged jets.\cite{Aaltonen:2011dr} 
Three observables are used to characterize the \ljets\ channel: the reconstructed invariant mass 
of the top quark for the two jet-permutations with the best and next-best fit-$\chi^2$ to the \ttbar\ hypothesis,
and the mass of the two untagged jets that provides an invariant mass ($m_{jj}$) closest to the world average value of
$M_W$. 
Two observables are used for the \dil\ channel: (i) the reconstructed $m_t$ based on the ``neutrino weighting'' algorithm
(described below), and (ii) the variable $m_{T2}$, which is related to the transverse mass of the decay remnants of the 
top-quark candidates.\cite{Lester:1999tx}
Two or three-dimensional templates for signal and background are constructed from MC samples using kernel-density 
estimators.\cite{Cranmer:2000du} The requisite distributions, generated at discrete input values of $m_t$, 
are smoothed and 
interpolated using the local polynomial-smoothing method of Ref.\cite{Loader}.
The resulting $m_t$ is $172.2 \pm 1.2 {\rm (stat)} \pm 0.9 {\rm (syst)}$~GeV in the \ljets\ channel,
and $m_t = 170.3 \pm 2.0 {\rm (stat)} \pm 3.1 {\rm (syst)}$~GeV in the \dil\ channel.\cite{Aaltonen:2011dr}
The same technique is exploited in an NN-based selection of \met+jets events that relies on the two reconstructed invariant masses
of the top quark and $m_{jj}$.
This yields: $m_t = 172.3 \pm 2.4 {\rm (stat)} \pm 1.0 {\rm (syst)}$~GeV in 5.7~\ifb\ of data.\cite{Aaltonen:2011kx}

CDF also uses the template method to measure $m_t$ in the alljets channel in 5.8~\ifb\ of data.\cite{Aaltonen:2011em}
Templates are formed for the mass reconstructed from the $b$-light-jet system and the $m_{jj}$ that correspond
to the best fit 
to a \ttbar\ hypothesis, following a NN-based selection.
The measured top-quark mass is $m_t = 172.5 \pm 1.4 {\rm (stat)} \pm 1.5 {\rm (syst)}$~GeV.

D0 also uses a template method for the dilepton analysis in 5.3~\ifb\ of data, to measure the top-quark mass based on a neutrino
weighting method\cite{Abazov:2012rp} that integrates over the rapidities assumed for the two neutrinos, using
the kinematic constraints of the \ttbar\ hypothesis that depend on $m_t$. 
Weights are assigned to each choice of rapidities by comparing the
resulting solutions for the summed $|\vec{p}_{T1}+\vec{p}_{T2}|$ of the two neutrinos to the measured value of \met. The mean 
and RMS values of the distributions in event weights are used as observables to extract the most probable mass of the top quark.
In this dilepton analysis, the dominant systematic uncertainties (from jet energy calibration) are reduced using a correction 
obtained from $\ttbar \to \ljets$ events.
D0 obtains $m_t = 174.0 \pm 2.4 {\rm (stat)} \pm 1.4 {\rm (syst)}$~GeV.\cite{Abazov:2012rp}

The ME method is based on using all measured kinematic quantities to construct a probability 
for each event that relies on a leading-order matrix element, and that integrates over the unmeasured quantities.
This method offers maximal statistical sensitivity to $m_t$, as it uses all the available kinematic information to 
weight events according to their degree of agreement with background or signal hypotheses. 
It is however a rather CPU intensive formulation.
This powerful technique was developed to measure the top-quark mass, and was adapted subsequently to measure 
the $W$ helicity in top-quark decays and to use as a discriminant in measuring the single top-quark cross section, 
\ttbar\ spin correlations, as well as to search for the Higgs boson at the Tevatron.
The event probability is constructed from signal and background probabilities, weighted by their fraction
of contributions.
The probability for signal is found by convoluting the parton-level differential cross section for 
$\qqbar \to \ttbar$ with parton distribution functions (PDF) and resolution functions that 
take account of detector resolution as well as parton evolution.
These transfer functions $W(x,y)$ correspond to the probability of observing a given measured quantity $x$ that 
corresponds to a parton level quantity $y$. When the analysis uses {\it in-situ} jet calibration, the 
jet transfer functions can be expressed in terms of an overall jet energy scale factor that is determined
simultaneously with $m_t$. 
The background probability is also defined in the analysis through an appropriate matrix element.
The likelihood function for a given event sample is obtained from the product of the individual event probabilities.

CDF and D0 have used the ME technique in analyses of \ljets\ and \dil\ channels.
CDF considers the $gg \to \ttbar$ matrix element in addition to the $\qqbar \to \ttbar$ process, and takes
account of both momentum and angular resolutions of jets in the transfer functions.
Signal is discriminated from background through a NN. Using a quasi-MC technique,
the ME is integrated over the dimensions representing the kinematics of the final state.
In 5.6~\ifb\ of data, CDF measures $m_t = 173.0 \pm 0.7 {\rm (stat)} \pm 1.1 {\rm (syst)}$~GeV.\cite{Aaltonen:2010yz}
For the signal probability, the D0 analysis relies only on the $\qqbar \to \ttbar$ matrix element, and the 
$W$+4~partons process for describing the background from $W$+jets and multijet events.
The result is $m_t = 174.9 \pm 0.8 {\rm (stat)} \pm 1.2 {\rm (syst)}$~GeV for 3.6~\ifb\ of data.\cite{Abazov:2011ck}
In the \dil\ channel, CDF uses event selections based on NN training using neuroevolution.\cite{Stanley:2002zz}
Background probabilities are contructed with matrix elements for the $Z/\gamma^*$+jets and $W$+jets matrix processes.
Using 2~\ifb\ of data, CDF obtains $m_t = 171.2 \pm 2.7 {\rm (stat)} \pm 2.9 {\rm (syst)}$~GeV.\cite{Aaltonen:2008bd}
D0 uses $Z$+2~jets matrix elements for the background probabilities in their \dil\ analysis.\cite{Abazov:2011fc}
An additional transfer function is used to describe the energy of the final state lepton for the
$Z \to \tau^+ \tau^-$ background in the $e\mu$ channel.
In 5.4~\ifb\ of data, D0 finds
$m_t = 174.0 \pm 1.8 {\rm (stat)} \pm 2.4 {\rm (syst)}$~GeV.

The third general way of extracting $m_t$ is referred to as the ideogram method, and can be thought of 
as an approximation to the ME approach. The procedure defines a probability for observing the reconstructed
$m_t$ in an event that is based on the mass resolution and an assumed input value of $m_t$.
Specifically, the probability for signal is obtained through a convolution of a Gaussian for the mass resolution 
with a Breit-Wigner characterizing the decay of top quarks, while the background
probability is taken from MC simulation.
D0 has performed a measurement using this technique in the \ljets\ channel by factorizing the probabilities 
for signal and background. With {\it in-situ} jet calibration, in 0.43 \ifb\ of data, D0 obtains
$m_t = 173.7 \pm 4.4 {\rm (stat+JES)} ^{+2.1}_{-2.0} {\rm (syst)}$~GeV.\cite{Abazov:2007rk}
CDF measures $m_t$ in the alljets channel using the same technique in 0.31 \ifb\ of data, and finds
$m_t =  177.1 \pm 4.9 {\rm (stat)} \pm 4.7 {\rm (syst)}$~GeV.\cite{Aaltonen:2006xc}

The above measurements in the \ljets\ and alljets channels are limited by systematic uncertainties.
For the full Tevatron data, this is also expected to be the case even for the \met+jets channel.
The largest systematic uncertainties in the \ljets\ analyses arise from the residual uncertainty on JES and 
from modeling of signal, for the alljets channel it is the 
modeling of signal and background and the \dil\ channel is limited by the uncertainty on JES.
However, as the statistical uncertainty decreases, the uncertainty from {\it in-situ} jet calibration also 
decreases with more data. 

Different measurements of $m_t$ at the Tevatron can also be combined to improve the overall uncertainty on $m_t$.
And in fact, eight measurements from CDF and four from D0 have been combined using the BLUE method to account for the systematic uncertainties of the input measurements 
and their correlations.
This yields a value of
$m_t = 173.18 \pm 0.56 {\rm (stat)} \pm 0.75 {\rm (syst)}$~GeV, 
which has a precision of 0.54\%.\cite{topmassworldaverage}
The combination has a $\chi^2$ of 8.3 for 11 degrees of freedom, corresponding to a 69\% probability for agreement
among the twelve input values.
The CDF and D0 \ljets\ measurements carry the largest weights in the combination, which is followed by the 
CDF measurement in the alljets final state. 

All the measurements presented so far rely on the reconstructed decay products of the top-quarks and assume that 
the measured top-quark mass is close to the ``pole'' mass. However, 
for a colored particle such as a top quark, this definition is intrinsically ambiguous by a value of the order of 
$\Lambda_{\rm QCD}$.\cite{Beneke:1994sw,Bigi:1994em,Smith:1996xz}
Apart from this theoretical ambiguity, additional questions of interpretation arise, as the experimental 
measurements
are calibrated using MC generators that include models for parton evolution (``showering'') and hadronization. These
processes also introduce an ambiguity of the order of $\Lambda_{\rm QCD}$ in interpreting $m_t$ as the pole 
mass.\cite{Buckley:2011ms}
An alternative method, mostly free of these ambiguities is based on extracting $m_t$ by comparing the observed
\ttbar\ cross section with theoretical predictions. 
Determining the mass from the cross section is less precise than using the direct methods described above, but
it provides $m_t$ in a well-defined renormalization scheme.
D0 has performed such an extraction using the measured \ttbar\ cross section in 
5.3~\ifb\ of \ljets\ data, for different assumptions of $m_t$.\cite{Abazov:2011pta}
The top-quark mass is extracted as the most probable value of a normalized joint-likelihood function formed
from the theoretical prediction with the measured cross section, taking account of the uncertainties from 
choices of PDF, renormalization and factorization
scales, and other experimental uncertainties. Using the NLO+NNLL calculation 
of Refs.\cite{Moch:2008qy,Langenfeld:2009wd}, D0 measures the pole mass to be $m_t = 167.5^{+5.4}_{-4.9}$~GeV.
This method is also used to extract the mass in the $\overline{MS}$ renormalization scheme.\cite{Abazov:2011pta}

Because the top quark decays before hadronizing, \ttbar\ events provide a unique opportunity to study the properties
of an essentially bare quark. Both template and matrix-element methods have been extended to check the 
CPT theorem, namely the conservation under the
product of the charge conjugation (C), parity conjugation (P) and time reversal (T) operation
in the top-quark sector. 
Tests have been performed by measuring the $t- \bar{t}$ mass difference.
CDF uses the template method in the \ljets\ channel with 8.7~\ifb\ to get 
$m_t - m_{\bar{t}} = -1.95 \pm 1.11 {\rm (stat)} \pm 0.59 {\rm (syst)}$~GeV.\cite{Aaltonen:2012zb}
D0 measures the mass difference using the matrix-element method in the \ljets\ channel with 3.6~\ifb\ of data
$m_t - m_{\bar{t}} = 0.8 \pm 1.8 {\rm (stat)} \pm 0.5 {\rm (syst)}$~GeV.\cite{Abazov:2011ch}
Both results agree within less than 2 standard deviations with the CPT-conserving hypothesis of no mass difference.

\subsection{Other Properties}

With the aim of revealing whether the massive quark observed is indeed the top quark postulated by the SM, other properties
besides its mass are analyzed at the Tevatron. One of the first of such measurements  was the study of 
the helicity of the $W$~boson produced in
the $t \to Wb$ decay. In the SM right-handed $W^+$ bosons are strongly suppressed by the V-A structure of the EW interaction. In particular, for $m_t=172.5$~GeV, the $W^+$  helicity is expected  to have a
longitudinal component of $f^0=0.696$ and a
left-handed component of $f^{-}=0.303$.  These helicity fractions obtained from a first-order in  perturbative 
expansion,\cite{wheltheory} can be  affected through higher-order EW effects or uncertainties on $m_t$, $m_W$ or $m_b$
by 1-2 \%.\cite{wheltheory2,wheltheory3,wheltheory4} Significant deviations from these predictions can indicate therefore the presence of new physics.  
First results on the observed helicities were obtained in the \ljets\ channel through studies of distributions of the helicity angle ($cos \theta^*$) which is the angle 
of the down-type fermion  (charged lepton) in the rest frame of the $W$ boson  relative to the direction of  the top quark  direction in the 
$t\bar{t}$ rest frame. This was done by comparing data with  templates extracted from MC simulations that were generated with different 
values of $f^+$ for the fixed (expected) value of $f^0$. These results were rapidly improved by the addition of the dilepton mode, 
and the use of results from the  $W \to q' \qbar$ decays. For the latter, the down-type quark jet is chosen at random in the 
calculation of  $cos\theta^*$, as this inclusion adds sensitivity to the measurement. More recent analyses 
include model-independent fits to simultaneous measurements of $f^0$ and $f^+$. Performing a joint binned-likelihood fit  to 5.4~\ifb\ in the \ljets\ and \dil\ decay modes, D0 finds
$f^0=0.669 \pm 0.078{\rm (stat)} \pm 0.065{\rm (syst)}$ and $f^+=0.023 \pm 0.041{\rm (stat)} \pm 0.034{\rm (syst)}$ 
for a model-independent fit, and, respectively, 
$f^+=0.010 \pm 0.022{\rm (stat)} \pm 0.030{\rm (syst)}$ and $f^0=0.708 \pm 0.44{\rm (stat)} \pm 0.048{\rm (syst)}$ 
for fits with $f^0$ or $f^+$ fixed to their SM values respectively.\cite{d0whel}  CDF performs this measurement using 2.7~\ifb\ of data in the \ljets\ channel, that 
introduces a likelihood technique based on a matrix elements for \ttbar\ production as well as for 
 the main background process from $W$+jets production.\cite{cdfwhel} This technique was developed for the top-quark mass 
measurement, and utilized instead an expression for the ME in terms of the $W$ boson helicity 
fractions and $cos\theta^*$. The study was recently updated using all the data, determining simultaneously $f^0=0.726 \pm 0.066{\rm (stat)} \pm 0.067{\rm (syst)}$ and $f^+=-0.045 \pm 0.044{\rm (stat)} \pm 0.058{\rm (syst)}$.\cite{cdfwhel2}
For the dilepton channel, the measurements are performed using a fit of the two-dimensional space of the measured $cos\theta^*$ 
in  a sample of 5.1~\ifb.\cite{cdfwheldil} Combining this with the above 2.7~\ifb\ \ljets\ result yields  $f^0=0.84 \pm 0.09{\rm (stat)} \pm 0.05{\rm (syst)}$ 
and  $f^+=-0.16 \pm 0.05{\rm (stat)} \pm 0.04{\rm (syst)}$ for the simultaneous measurements and 
$f^+=-0.07 \pm 0.02{\rm (stat)} \pm 0.04{\rm (syst)}$ and $f^0=0.64 \pm 0.06{\rm (stat)} \pm 0.05{\rm (syst)}$
when  $f^0$ or $f^+$  are fixed, respectively, to their SM expectations.
Results from both experiments are combined to yield $f^0=0.722 \pm 0.081{\rm (stat+syst)}$ and  
$f^+=-0.033 \pm 0.046{\rm (stat+syst)} $ and when  fixing one of the helicity fractions to the SM prediction the results are
$f^0=0.682 \pm 0.057{\rm (stat+syst)}$ and $f^+=-0.015 \pm 0.035{\rm (stat+syst)}$.\cite{whelcombo} 
These measurements are consistent with the SM with no indication of the presence of contributions from new phenomena. 

Studies have also been related to properties of the $\ttbar$ production process.
A measurement of the fraction of $gg$ fusion process ($f_{gg}$) in the \ttbar\ production performed at CDF 
exploited the difference in kinematic characteristics of $gg$ and $\qqbar$ contributions to distinguish the two mechanisms. Eight variables, describing production and 
decay properties, all sensitive to the production mechanism, are fed into a NN for two b-tagged event categories: 1 and $>1$  $b$-tagged jets. The outputs of the NN are formed  into templates to represent  background, $\qqbar$, and $gg$ events that are used in a likelihood function that is maximized to find the estimator for $f_{gg}$. 
Using the Feldman-Cousins prescription,\cite{Feldman:1997qc} measured values are mapped to a range of MC-generated true fractions. 
For the \ljets\ channel, in a sample of  1~\ifb\ of data CDF finds $f_{gg} < 0.33$ at a 68\% CL.\cite{cdfgg} 
This result is combined with 
another measurement that relies on the higher probability for a primary gluon, than for a quark,  to radiate a 
low energy gluon in the production process, obtaining
 a value of $f_{gg} = 0.07 ^{+0.15}_{-0.07}$, in agreement with the SM prediction.\cite{cdfgg}

While at leading order, QCD predicts that angular distributions of $t$ and $\tbar$ production should be forward-backward symmetric at the Tevatron, a 
positive asymmetry (more $t$ and $\tbar$ produced along the incident $p$ and $\pbar$), $A_{FB}$, is expected at higher orders.\cite{theoryafb1,theoryafb2,theoryafb3} Negative contributions to $A_{FB}$ arise from the interference of diagrams with initial and final state radiation, 
while positive terms arise from interference of the Born and box diagrams in two body $\ttbar$ production. 
D0 explores $A_{FB}$ defined in terms of the rapidity difference ($\Delta y$) between the top and antitop quarks in 
\ljets\ events in a sample of 5.4~\ifb. After correcting for acceptance and detector resolution through an unfolding 
method with fine binning and explicit regularization, D0 finds
 $A_{FB} = (19.6 \pm 6.5)\%,$\cite{d0afb} to be compared with the 
prediction from the Monte Carlo generator MC@NLO of $(5.0 \pm 0.1)\%$.\cite{d0afbMCNLO}
An alternative approach is also performed by calculating the asymmetry 
based on the rapidity of the lepton. This result, which does not depend on the full reconstruction of the \ttbar\ 
system, yields $A^{l}_{FB} = (15.2 \pm 4.0)\%$\cite{d0afb} which is predicted to be  $A^l_{FB} = (2.1 \pm 0.1)\%$.\cite{d0afbMCNLO}
These values disagree with 
 expectations by up to 3 standard deviations. However, no statistically significant dependence of $A_{FB}$ is observed on the 
invariant mass of the \ttbar\ system ($m_{\ttbar}$)
or on $|\Delta y|$.
CDF, using an integrated luminosity of 9.4~\ifb\ to perform an inclusive measurement of $A_{FB}$, and to examine the dependence on 
 kinematic properties in the \ljets\ channels, corrects the reconstructed $A_{FB}$ for 
acceptance and resolution of the detector by using a regularized  algorithm, to unfold the resolution,  which is based on Singular Value 
Decomposition,\cite{cdfafbSVD,cdfafbSVD2} 
and bin-by-bin correction for acceptance obtained from the Monte Carlo generator POWHEG.\cite{cdfafbpowheg}
The inclusive result is $A_{FB}=(16.4 \pm 4.5)\%,$\cite{cdfafb} 
which exceeds the NLO prediction from POWHEG by 2 standard deviations (including a 30\% uncertainty on the prediction,\cite{cdfafbtheory} and EW corrections that amount to a factor of $\approx 26\%)$ \cite{cdfafbtheory2,cdfafbtheory3,cdfafbtheory4}. 
A linear fit is carried out as a function of $m_{\ttbar}$, finding a slope of $(15.2 \pm 5.0) \times 10^{-4}$ GeV$^{-1}$, which 
exceeds by 2.3 standard deviations the NLO prediction of $(3.4 \pm 1.2) \times 10^{-4}$~GeV$^{-1}$. A fit 
to $|\Delta y|$ yields
a slope of $(28.6 \pm 8.5) \times 10^{-2}$, which is 2.1 standard deviations away from the 
expectation of $(10.0 \pm 2.3) \times 10^{-2}$. 
The significance of the observed difference between data and theory for reconstructed $\ttbar$ events, after background subtraction, is reflected by the p-values  for the slopes to have fluctuated to values as large as observed in data. These probabilities are
$14.7 \times 10^{-3}$ for  $|\Delta y|$ and $7.4 \times 10^{-3}$ for $m_{\ttbar}$, corresponding to 2.2 standard deviations 
and 2.4 standard deviations, respectively. These results have stimulated many new theoretical work, not only in the SM context but also new physics models that would explain the  observed asymmetry and that should accommodate the consistency with the SM of the measured cross section and $m_{\ttbar}$ spectrum.\cite{afbtheoreview} In addition, many cross-checks were performed on these results by both CDF and D0, including a check of  the 
asymmetry as function of the transverse momentum of the \ttbar\ system.
Such measurements contribute to the on-going studies needed to establish the origin of these discrepancies. In fact  both collaborations are also exploring the issue of the asymmetry  in other 
final states, such as the dilepton channel. In this mode, D0 measures angular asymmetries based on $\eta$ distributions of charged 
leptons. The resulting $A^{l}_{FB}$ is $5.8 \pm 5.1 {\rm (stat)} \pm 1.3 {\rm (syst)}\%,$\cite{d0dilafbl} 
which, including QCD and EW corrections,  is in agreement with the 
 MC@NLO  prediction of $(4.7 \pm 0.1)\%$.\cite{d0dilafbltheory}

\subsection{Properties Made Accessible through Large Statistics}

While the data sample of Run I  was too small to
study all aspects of the properties of top quarks, the large amount of data
collected by now opens up the possibility of studying more subtle 
properties. In particular, the charge of the top quark, and
$t\bar{t}$ spin correlations have now turned into precision measurement,
 while effects of color flow from the presence of   $W$~bosons in $t\bar{t}$ events (but not in background)
and $t\bar{t}$ production associated with a photon were only recently considered for study.  

While the charge of the top quark in the SM  is predicted to be +2/3 of
the electron charge, an exotic charge of $-4$/3 could also be possible.\cite{Chang:1998pt}
CDF and D0 both perform measurements of the charge of the top quark. The
measurements are performed in the \ljets\ final state, in events with
at least one $b$-tagged jet. A kinematic fit is performed to assign
the final state $t$ and $\bar{t}$ decay products to their proper top
and antitop quark; where constraints from $m_W$ and $m_t$ are applied in the analysis. The  measurements of charge rely on the
observed charge of the lepton from $W \to l\nu$
decay, combined  with the charge of the $b$-jet from the same or the other top
quark. D0 performed the
first  measurement of the charge of the top quark using 0.37~fb$^{-1}$ of data, where a jet
charge algorithm was applied to extract the charge of the $b$-jet.\cite{d0_charge} In this method, a weighted sum of the charges of the
tracks is calculated within the jet and calibrated using an orthogonal
data sample enriched in $b\bar{b}$ events, where one of the $b$-jets
is also tagged through the presence of a soft (low-\pt) muon. This calibration is used to analyzed the $t\bar{t}$ sample, providing templates representing  SM and exotic choices to model the 
charge of top quarks. The result of a fit to data excludes in the exotic hypothesis
 at a 92\% CL. 
 
CDF exploits an alternative approach, requiring at least one jet 
to be $b$-tagged through a displaced vertex,  and at least one (which can
be the same jet) to contain a soft lepton from semileptonic $B$
decay. Using 2.7~fb$^{-1}$ of data, CDF excludes the exotic model
at a 95\% CL.\cite{cdf_alternativecharge}

Another way to access the charge of the top quark is to study directly the electromagnetic
coupling strength in top-quark electromagnetic interactions through photon radiation in \ttbar\ events.
The $\ttbar \gamma$ coupling parameters are also sensitive to new physics models.\cite{Baur:2001si}
CDF performed a measurement of the cross section for $\ttbar \gamma$ production together with the inclusive
production of \ttbar\ events using a selection optimized for the $\ttbar \gamma$ candidates
in the \ljets\ and \dil\ samples and requiring at least one jet to be identified as coming from a $b$~quark.
The $\ttbar \gamma$ sample requires the photon to have $E_T>10$~GeV and to be in the central region of the detector. 
In the $\ttbar \gamma$ signature, the background is dominated by events in which an electron is misidentified as a photon.
Using data corresponding to 6~\ifb, 30 $\ttbar \gamma$ candidates are observed compatible with the predictions from the SM.
The $\ttbar \gamma$ cross section yields $0.18 \pm 0.08$~pb and the ratio of production of $\ttbar \gamma$ to
\ttbar\ is $0.024 \pm 0.009$.
This corresponds to the first experimental evidence for $\ttbar \gamma$ production.\cite{Aaltonen:2011sp}

Another analysis that gained sensitivity with 
collected data is the study of $t\bar{t}$ spin correlations. Despite that
the top quark is expected to be produced unpolarized at lowest order in the SM, the spin
of the $t$ and $\tbar$ quarks are predicted to be correlated. The short lifetime of the top quark assures that the information about its spin is
preserved in its decay products, which can be used to  measure the
 spin correlations of $t$ and $\tbar$. While previous data samples
were not sufficiently sensitive to provide definitive results on spin correlations, several studies  performed more recently at CDF and D0 in the dilepton
and the \ljets\ channels have been more conclusive. 

 CDF and D0 use template-based
methods that rely on the fact that the double differential cross
section, $1/\sigma \times d^2
\sigma /(d \cos \theta_1 d \cos \theta_2)$, can be written as $1/4
\times 
(1-C \cos \theta_1 \cos \theta_2)$, where $C$ is the spin correlation
strength, and $\theta_1$ and $\theta_2$ are respectively, the angle of the down-type
fermions from $t \to W^{+}b$ and $\tbar \to W^{-}b$ decays of the $W^{+}$ and $W^{-}$ bosons in the  $t$ or $\tbar$ quark rest frame relative to some chosen  quantization axis. 
The SM prediction for the spin correlation strength ($C$)
depends on the collision energy ($\sqrt{s}$) and the choice of quantization axis,
and at NLO corresponds to  $C=0.78$ for $\sqrt{s}=1.96$TeV, as defined relative to the beam direction.\cite{bernreutherspin}
The optimal choice for final-state
particles is  the charged lepton and the down-type quark from the $W$-boson
decay, both with spin-analyzing power of unity. Because of the 
experimental challenge of distinguishing up-type from down-type quarks,
the dilepton channel has greatest sensitivity to spin
correlations. The D0 experiment extracts $C$ by forming  templates for $C=0$ and for the values of  $C$ expected for the coefficient  of $\cos
\theta_1 \cos \theta_2$ in the SM, and fitting  these two possibilities  in an analysis of 5.4~fb$^{-1}$ of data.
In the beam basis, $C$ is found to be
$C=0.10 \pm 0.45 {\rm (stat+syst)}$, in agreement
with the prediction of the SM.\cite{d0_dilepspin}
 The first measurement of $C$ by CDF  is in the  \ljets\ channel, where  templates of same and opposite $t\bar{t}$
helicity are fitted to the data. Using 4.3~fb$^{-1}$ of data, CDF measures $C=0.72 \pm 0.64
{\rm (stat)} \pm 0.26 {\rm (syst)}$ in the beam basis.\cite{cdf_ljetsspin}

In addition to the template based method, D0 explores a technique
based on calculating matrix elements that consider spin correlations (labeled as $H=c$) and matrix elements with uncorrelated spins
($H=u$). From these matrix elements,  a discriminant $R$ can be constructed as $R=P_{sig}(H=c)/[P_{sig}(H=c)+P_{sig}(H=u)]$.\cite{melnikovschulze}
Using the same data sample of 5.4~fb$^{-1}$ as used in the template-based analysis in
dilepton events, the method provides a 30\% improvement in sensitivity,
yielding $C=0.57 \pm 0.31 {\rm (stat+syst)}$.\cite{d0_dilepmespin} Applying the ME-based method to 5.3~fb$^{-1}$ of \ljets\ events, and combining the result
with the
measurement in the dilepton final state, yields first evidence for a
non-vanishing $t\bar{t}$ spin correlation.\cite{d0_ljetsdilepmespin}

D0 performed the first study of color flow
in $t\bar{t}$ events. Since color charge is a conserved quantity in QCD,
two final-state particles on the same line of color flow are termed to be
color-connected to each other. Using 5.3~fb$^{-1}$ of \ljets\ events,
D0 exploits a tool called jet pull,\cite{theorypull} which is based on the measurement
pattern of jet energy  distributed in the $\eta$-$\phi$ plane, and measures the
color flow between a pair of jets, in an attempt  to distinguish color-octet
from color-singlet states. For a color-singlet state, the pulls
of both jets tend towards each other, in contrast to a jet pair from a
color-octet state, where the pulls point in opposite directions along the
beam axis. The known environment of \ljets\ $t\bar{t}$ events provides
a testing ground for this tool, before it can also be applied to searches
for BSM contributions. The two light jets from the decay of the $W$ boson
are expected to originate from a color singlet. By introducing a
hypothetical ``$W$'' boson that decays as color octet, and
comparing templates of  jet pull for octet and
singlet components to data, the fraction of color-singlet
decays is found to be  $f = 0.56 \pm 0.38{\rm (stat + syst)} \pm
0.19{\rm(MCstat)}$, with an expected exclusion
of a color-octet ``$W$'' boson at the 99\% CL.\cite{d0_colorflow}

\subsection{Properties extracted from multiple inputs}


Certain top-quark properties, or searches for new physics in the top-quark sector,
can be elucidated by combining two a priori independent measurements that can yield
 additional information  on the parameters of the top quark.

An example of a combination of different measured parameters that yields a new result  is 
the determination of the width of the top quark ($\Gamma_t $).
In the SM, $\Gamma_t$ can be computed from the value of $m_t$. For $m_t=172.5$~GeV, the expectation 
 is $\Gamma_t = 1.33$~GeV. The value of $\Gamma_t $ can be affected by the presence of new physics. 
With 4.3~\ifb\ of data, CDF extracts $\Gamma_t $ in the \ljets\ channel directly 
using a standard template method based on the distribution of the reconstructed $m_t$ values,
with $\Gamma_t $ as the parameter of interest.
An upper limit of $\Gamma_t < 7.6$~GeV is established at 95\% confidence by
applying a Feldman-Cousins approach.\cite{Aaltonen:2010ea}
The resolution of the reconstructed top-quark mass is limited especially  by the resolution in JES 
and by the uncertainty on modeling the $m_t$ resonant spectrum.	
Hence, since the predicted $\Gamma_t $ is far smaller than the mass resolution of the reconstructed top quark,
it is very difficult to measure $\Gamma_t $ directly.
To overcome this, D0 extracts $\Gamma_t$ from its partial width $\Gamma (t \to Wb)$, which is   
determined from the t-channel of the  single top-quark production cross section.\cite{Abazov:2011rz} 
This coupled with  the top-quark branching fraction ${\cal B}(t \to Wb)$ measured using the ratio 
$R = {\cal B}(t \to Wb) / {\cal B}(t \to Wq)$\cite{Abazov:2011zk}, and the assumption  that
$\Gamma_t = \Gamma (t \to Wb) / {\cal B}(t \to Wb)$, (namely  that 
${\cal B}(t \to Wq) = 1$) and that the $Wtb$ coupling is the same in the production and decay 
of the top quark, provides a measure of $\Gamma_t $.
Using the above inputs from analyses using 5.4~\ifb\ of data, and applying Bayesian
techniques to combine the measurements, D0 obtains
$\Gamma_t = 2.00^{+0.47}_{-0.43}$~GeV for $m_t=172.5$~GeV, in agreement with the SM.\cite{Abazov:2012vd}

Again, based on multiple inputs, D0 searches for contributions from anomalous  top-quark
couplings in 5.4~\ifb\ of data\cite{Abazov:2012iwa}, and, in particular for right-handed vector couplings ($f^R_V$) or 
left or right-handed tensor couplings ($f^L_T, f^R_T$), in addition to the $V-A$ left-handed 
$f^L_V$ interaction of the SM.
This search is performed by combining a measurement of the $W$ helicity in the \ljets\ and \dil\
channel\cite{Abazov:2010jn} with the measurement of the single top-quark cross section.\cite{Abazov:2011pm}
The limits are obtained by setting all but  one of the  anomalous coupling to zero.
$W$ helicity  is especially sensitive to $f^R_V$, while the single top-quark
cross section is in particularly sensitive to $f^R_V$ and $f^L_T$.
Using a Bayesian statistical analysis to combine the measurements, D0 sets the following limits:
$|f^R_V|^2<0.30$, $|f^L_T|^2<0.05$ and $|f^R_T|^2<0.12$, where no assumptions on $|f^L_V|$ are made.

\section{Searches in the Top Sector}
\label{sec:search}


Besides the precise understanding of the production and properties of the top quark,
where deviations from the SM prediction could indicate physics beyond
the SM, a variety of direct searches has also been performed for new signals
in the top-quark sector.
A broad spectrum of
search methods has been developed and applied starting with classical
searches for resonant peaks, to more elaborate methods, such as the combination of information
from different final states, or the use of
multivariate discriminants. In this section, we give a brief overview
of the most recent direct searches in the top-quark sector.  

\subsection{Classical Searches}
The most common way of searching for a new particle is to look
for resonant peaks above a known background in the distribution of a specific variable, such as
the \ttbar\ invariant mass. Searches of this kind have been carried out
in the $t\bar{t}$ mass spectrum,
as well as for fourth-generation $b^{'}$ or
$t^{'}$ quarks, and for the stop quark, supersymmetric partner of the top quark.

No $t\bar{t}$ resonances are expected in the SM, but many BSM models
such as topcolor assisted technicolor models,\cite{ttbarreso_technicolor}
predict such resonance. 
Both CDF and D0 have searched for a narrow
resonance $X$, assuming  a width of $\Gamma_X=1.2\% M_X$, which is
smaller than the resolution of the detector. Using events in the \ljets\ final
state, the searches were carried out in spectra that reflect the value of $m_{t\bar{t}}$.
The mass variable can be defined either through some 
kinematic considerations that contain partial
information about the escaping neutrino, or by using
a kinematic fitter that constrains the jets, the charged lepton and
the \met\ to the known features of the $t\bar{t}$
hypothesis, and adjusts the energies of the jets within their
resolution. The most recent searches for a  $t\bar{t}$ resonance
extract limits on $\sigma(p\bar{p}\rightarrow X)\times B(X\rightarrow t\bar{t})$ as a function of 
$M_{X}$ in events with at least three jets in
9.45~fb$^{-1}$ of data at CDF and 5.4~fb$^{-1}$
at D0. By considering the benchmark model of  topcolor-assisted technicolor, a $Z^{'}$ with $m_{Z'}<835$~GeV is
excluded by D0,\cite{ttbarreso_d0} and $m_{Z'}<915$~GeV by
CDF,\cite{ttbarreso_cdf} both at 95\% CL. In addition, D0 shows that the limits do not depend on the
couplings of the $t\bar{t}$ resonances, i. e. whether they are purely axial-vector (A), vector (V), or SM-like (V-A). 
CDF also considers alljet events, where the reconstruction of the
correct invariant mass from a resonant contribution is diluted by the large number
of possible jet combinations. CDF uses the ME approach to calculate
per-event probability densities to minimize the impact of the large background, and,
following event selection based on a NN, searches for $t\bar{t}$ resonances in $2.8$~fb$^{-1}$ of
data, resulting in an exclusion of $Z^{'}$
with $m_{Z'}<805$~GeV at 95\% CL.\cite{cdf_alljets_ttbarreso} CDF performs another
search for $t\bar{t}$ resonances in a search for a massive vector color-octet boson
(e.g., a massive gluon)  in the \ljets\ final state in 1.9~fb$^{-1}$ of data, setting limits on the 
coupling strength for different masses $m$ and $\Gamma/m$ ratios.\cite{massivegluon}

Exploring  models that can accommodate the anomalous $A_{FB}$ results, 
CDF has performed a first search for top-quark+jets resonances, seeking signs for the production of a new heavy 
particle (decaying into $\bar{t}q$) in association with a top quark. In $8.7$~fb$^{-1}$ of data, the study selects 
events with one lepton, \met\ and at least five jets and reconstructs the mass of the $tj$ system. 
Finding data to be consistent with the SM expectations, 95\% CL upper limits are determined on the production 
cross section for different possible masses of the new particle.\cite{topjetreso_cdf}

Because the extension of the SM to a 
fourth generation, with  massive up-type  $t^{'}$ and down-type
 $b^{'}$ quarks remains a distinct possibility, both collaborations have searched for $t^{'}$
 pair production, assuming  $t^{'} \rightarrow Wq$ decays.  CDF
performs parallel searches for all the down quarks in  $t^{'}$ decays ($d$, $s$ or $b$), while D0 assumes just a $b$-quark as a possibility. Since the $t^{'}$ is expected to
have a mass $m_{t^{'}} > m_t$,  the search strategy is based on
looking for events with larger fitted $Wb$ masses  and a larger
scalar sum of the $p_T$ values of the lepton and jets than expected in SM $t\bar{t}$ decays. 
The search is performed in the
$\ell$+jets final state, containing   at least four jets, of which, for D0,  at least
one is an identified $b$-jet candidate. The latest upper limits on 
  $\sigma(p\bar{p}\rightarrow t^{'}\bar{t}^{'})$ at the 95\% CL. as
function of $m_{t^{'}}$, are extracted using
5.3~fb$^{-1}$ of data at D0~\cite{d0_tprime} and 5.6~fb$^{-1}$ at CDF,\cite{cdf_tprime} resulting in  $m_{t^{'}}>285$~GeV and
$m_{t^{'}}>358$~GeV for $t^{'} \rightarrow Wb$ decays, respectively, and $m_{t^{'}}>340$~GeV for $t^{'}$ decays into a $W$ boson and any SM down quark for 
CDF.
CDF also searches for a massive fourth-generation $b^{'}$
quark, decaying into a $W$-boson and a top quark. Again, using \ljets\ events, considering 
the scalar sum of just the jet-$p_T$ values, which is sensitive to a $b^{'}$ signal in events with 
jets of large $p_T$ 
and large jet multiplicity,
 CDF, using 4.8~fb$^{-1}$ of data,  excludes a $b^{'}$ with mass $m_{b^{'}}<372$~GeV at 95\% confidence.\cite{cdf_bprime} 

The SM particles describe only  $\approx 4$\% of the energy content of the universe and the  rest consists of dark energy and 
dark matter. Possible candidates for dark matter (DM) could be long lived, weakly interacting massive particles (WIMPs),\cite{Turner:1989be} 
for example, such as the lightest supersymmetric particle, the neutralino. Recently, CDF performed a search 
for dark matter in the top-quark sector. CDF investigates  pair production of some unknown partner of the top quark, $T$, 
where $T$  decays into a top quark and a stable, neutral, weakly interacting particle ($A_0$). 
The search strategy relies on using the $t\bar{t}$ signature with large \met\ and a 
large transverse mass of the lepton and \met\ system in \ljets\ events. No deviations from the SM 
are observed by CDF, and upper limits are reported  on $\sigma(p\bar{p} \rightarrow T \bar{T})\times B(T\bar{T} \rightarrow tA_0\bar{t}A_0)$.\cite{cdfdmsearch} 
CDF also searches for dark matter in the $t\bar{t}$ final state in \ljets\ events, where 
the dark matter is produced through an unknown fourth generation $T^{'}$ quark that decays into a top quark and a 
dark matter candidate $X$. Using 4.8~fb$^{-1}$ of data, CDF excludes $T^{'}$ masses $m_{T^{'}}<360$~GeV 
for masses of $X<100$~GeV.\cite{cdfttbardm} A search for the same process is also performed in the 
alljets channel using 5.7~fb$^{-1}$ of data, increasing thereby the range of exclusion 
 to $m_{T^{'}}<400$~GeV for masses of $m_X < 70$~GeV.\cite{cdfttbardm_allhad} 
The analysis strategy focuses again on the spectrum in transverse mass 
 of the leptonically decaying $W$ boson, that acquires a broader spread in the signal due to the large \met\ contributed by  the 
dark-matter candidate in the \ljets\ channel. The value of \met\ divided by the square 
root of the total observed energy is used in the analogous analysis of the alljets final state. Using a similar strategy, a first search 
for production of  DM in association with a single top quark at hadron colliders was  performed at CDF using 7.7~fb$^{-1}$ of data.\cite{cdfmonotop} 
Utilizing the production mode $t+DM\rightarrow Wb+DM$, with the W boson decaying exclusively into $q^{'}\bar{q}$, the \met\ 
corresponds to the $p_T$ carried away by the DM particle. Finding the data consistent with  SM expectations, 
limits are determined on the production cross section as function of the mass of the dark matter candidate. 

Supersymmetric extension of the SM predicts the existence of a scalar partner of the
top quark, the stop quark ($\widetilde{t}$). CDF and D0 search for the
production of a $\widetilde{t}\widetilde{\bar{t}}$ quark pair, where each stop decays into a $b$~quark and
a chargino ($\chi_1$), and the chargino into a $W$-boson and a neutralino ($\chi_0$). The
neutralinos leave the detector without interacting, giving rise to
larger \met\ in the stop-pair signal than expected for
$t\bar{t}$ events. CDF searches for such events in the dilepton final
state using 2.7~pb$^{-1}$,\cite{cdf_stop} where the $m_{\widetilde{t}}$ is reconstructed to
discriminate signal from SM background. D0 searches in the \ljets\
final state using 0.9~fb$^{-1}$,\cite{d0_stop} where a
multivariate discriminant is defined on the basis of several variables.
Neither  search shows any signal, and
limits are set on the $m_{\widetilde{t}}$ for different choices of $m_{\chi_1}$.

\subsection{More elaborate Methods}
As in the example of the search for $\widetilde{t}\widetilde{\bar{t}}$ production, where the
 strategy changed from a direct search for a resonant peak to a more 
indirect method,  other search strategies  have also been
developed for BSM searches, such as searches for  $W^{'}$ bosons, flavor-changing
neutral currents (FCNC), and charged Higgs bosons.

Many models beyond the SM contain additional charged $W^{'}$ bosons. Searches in the single top-quark channel are performed 
for $W^{'}\rightarrow tb$ decays,  with both left and right-handed coupling to
fermions. The first search at D0,\cite{d0_wprime_old} using 0.9~fb$^{-1}$,  and  at CDF,\cite{cdf_wprime} using 1.9~fb$^{-1}$, focussed on finding a  $W^{'}$ with SM-like couplings, looking
for a peak in the invariant mass spectrum of the decay
products. A more recent search by D0 in 2.3~fb$^{-1}$ of data explores a
multivariate analysis, where several variables
are combined to form a discriminant, with the $W^{'}$ signal  trained
relative to backgrounds from the SM, assuming different couplings of the $W^{'}$
to the fermions.\cite{d0_wprime}

No contributions from FCNC are expected at lowest level in the SM. Consequently, observing such effect 
 would indicate the presence of physics beyond the SM. Both CDF and D0 searched for
effects from FCNC in the top-quark sector. At CDF, a search for $t\rightarrow Zq$ decays was performed in $t\bar{t}$ events
in final states that contain two leptons from the decay of the $Z$-boson
and a $t\rightarrow Wb \rightarrow q^{'}\bar{q}$ decay.  After splitting the sample into
 subsamples according to their content of $b$-jets, a 
$\chi^2$ variable is constructed to search for a peak in the $\ell^{+}\ell^{-}$ mass spectrum
in 1.9~fb$^{-1}$.\cite{cdf_fcnc} At D0, the search
for FCNC in $t\bar{t}$ events uses 4.1~fb$^{-1}$ of data, in 
events with three leptons in the final state, one from the
decay of the $W$-boson and two from the  $Z\rightarrow \ell^{+}\ell^{-}$ decay. For this
search, an excess is sought in the spectrum of the nominal reconstructed top mass, as well as in  the scalar
sum of the $p_T$ of the decay remnants of the candidate top quarks, resulting in
limits of $B(t\rightarrow Zq) < 3.2\%$ at 95\% CL.\cite{d0_fcnc} While the searches
for FCNC in $t\bar{t}$ events relies mostly on one or two variables, D0 also performed a search for the decay $t \rightarrow gu$
and $t \rightarrow gc$ in single top-quark events using a multivariate
discriminant. Exploring 2.3~fb$^{-1}$ of data, the dedicated training
of FCNC signal relative to the SM background yields limits of $B(t \rightarrow
gu)<2.0 \times 10^{-4}$ and  $B(t \rightarrow gc)<3.9 \times 10^{-3}$ at 95\% CL.\cite{d0_singletopfcnc} 
Using 2.2~fb$^{-1}$, CDF also searches for FCNC in single top-quark events, 
resulting in $B(t \rightarrow gu)<3.9 \times 10^{-4}$ and  $B(t \rightarrow gc)<5.7 \times 10^{-3}$ at 
95\% CL.\cite{cdf_singletopfcnc}

While decays of the top quark  in the SM almost always happen into a $W$-boson 
and a $b$~quark, other  models predict the extension of the Higgs
sector by at least one doublet. In these models a charged Higgs boson ($H^{\pm}$) that is 
lighter than the top quark is expected to exist, giving rise thereby to  the possibility of
$t\rightarrow H^{+}b$ decays. Both collaborations
perform a variety of searches for light charged Higgs bosons,
that consider $H^{+}$ decays into $\tau \nu$, $c \bar{s}$ or a CP-odd
neutral Higgs boson ($A$). Assuming pure $H^{+}\rightarrow c \bar{s}$
decays, CDF searches in the \ljets\ channel in 2.2~fb$^{-1}$ of data for a 
peak  in the invariant mass of the jets without $b$-tags.\cite{cdf_lightHplus} CDF also considers the
possibility of $H^{+} \rightarrow W^{+}A$ decay, searching for a 
deviation in the distribution of isolated tracks of low $p_T$ from the
decay of the $\tau$ leptons in 2.7~fb$^{-1}$.\cite{cdf_cpoddhplus} Another
search strategy for light charged Higgs bosons relies on the expected  differences in 
distributions of events in the different classes of  final states, resulting from different branching fractions  compared to predictions from just the
SM. At D0, a comparison of the number of events in the \ljets, dilepton,
and $\tau$+lepton final states is performed using 1.0~fb$^{-1}$ of data, assuming $B(H^{+}
\rightarrow \tau \nu)+B(H^{+} \rightarrow
c\bar{s})=100\%$.\cite{d0_lighthplus} A similar search is available from  CDF for 0.2~fb$^{-1}$ of 
data.\cite{cdf_oldhplus} None of the searches show a
deviation from the SM, and upper limits on $B(t \rightarrow H^{+}b)$
are set as function of $m_{H^{+}}$. At D0,  the
possibility of heavy $H^{+}$ bosons  is explored through 
 $H^{+} \rightarrow t\bar{b}$ decays. Using 0.9~fb$^{-1}$ of data, a
multivariate discriminant is trained for $H^{+}$ signal relative to 
SM background in single top-quark events, resulting in upper limits on
$\sigma(H^{+} \rightarrow t\bar{b})$ for several 2HDM scenarios.\cite{d0_heavyhplus}

In Section \ref{masssec}, the measurement of the difference in masses of the top and anti-top quark has been described, 
which is a test of the CPT theorem. 
D0 has recently explored the possibility of the violation of Lorentz
invariance  in the top-quark sector.
The issue of CPT invariance is related to Lorentz invariance, as violation of Lorentz invariance leads to the violation of 
CPT in particle interactions.\cite{Greenberg:2002uu}
For this search, the
time stamp is extracted for each luminosity block of recorded data to search for a dependence of the $t\bar{t}$ cross section in the \ljets\
final state on sidereal time  in 5.3~fb$^{-1}$ of data. No indication for
a time dependent   $\sigma_{t\bar{t}}$ is observed, resulting in
the first constraints on the standard-model extensions for violation of  Lorentz
invariance  in the top-quark sector.\cite{d0_liv}

\section{Summary}
Since the discovery of the top quark in 1995 by the CDF and D0 collaborations, a wide range of measurements and searches in the area of top-quark physics have been
carried out. By now, most of the available Tevatron data  has been analysed, providing strong evidence that the top quark is indeed the particle  expected in the SM. 
Measurements such as the production angular asymmetry of $t$ and $\bar{t}$, $t\bar{t}$ spin correlations, the mass of the top quark  and the production cross sections represent an important legacy of the Tevatron.
The $p \bar{p}$ initial state makes some of the measurements at the Tevatron unique,
 and complementary to what can be learned from $pp$ collisions at the LHC.

\section*{Acknowledgments}
We would like to thank our colleagues from the CDF and D0 collaborations for the hard work on analyses and detector operations, 
without which this review would not have been possible. In particular we would like to thank the 
top-quark physics groups and the CDF and D0 top-quark physics group conveners.  
We also would like to thank the Fermilab accelerator devision that enabled the successful run of the Tevatron. 
We gratefully acknowledge Tom Ferbel, who suffered through our non-native english.
Y.P. would like to acknowledge the support from the Helmholtz association.

\end{document}